\documentstyle[psfig]{./prhep97}

\def\Journal#1#2#3#4{{#1} #2 (#4) #3}

\def\NPA{{Nucl. Phys.} A}
\def\NPB{{Nucl. Phys.} B}

\def\PRC{{Phys. Rev.} C}
\def\PRD{{Phys. Rev.} D}

\newfam\BMath
\font\BMathL=cmmib10 
\font\BMathl=cmmib7
\font\BMathm=cmmib5
\textfont\BMath=\BMathL \scriptfont\BMath=\BMathl
\scriptscriptfont\BMath=\BMathm

\def\a{\alpha}

\def\j{\psi}

\def\be{\begin{equation}}
\def\ee{\end{equation}}
\def\bea{\begin{eqnarray}}
\def\eea{\end{eqnarray}}
\def\bfig{\begin{figure}}
\def\efig{\end{figure}}

\def\fref#1{Fig.~\ref{#1}}
\def\bfi{\begin{figure}}
\def\efi{\end{figure}}

\newcommand{\ncom}{\newcommand}
\ncom{\lan}{\langle}
\ncom{\ran}{\rangle}
\ncom\fx{\!\!\!\!}

\begin{document}



\authorrunning{S.M.H. Wong}
\titlerunning{{\talknumber}: Strong Coupling Improved 
Equilibration in High Energy Nuclear ...}
 

\def\talknumber{607} 

\title{{\talknumber}: Strong Coupling Improved Equilibration 
In High Energy Nuclear Collisions}
\author{S.M.H. Wong
(wong@theorie.physik.uni-wuppertal.de)}
\institute{Fachbereich Physik, Universit\"at Wuppertal, 
D-42097 Wuppertal, Germany}

\maketitle

\begin{abstract}
In high energy nuclear collisions, most calculations on the
early equilibration of the parton plasma showed that
the system does not come close to full equilibrium, especially
for the fermion components. However, since the system is
constantly evolving, the cooling effect due to expansion
will lead to a decrease of the average parton energies. 
So the interactions should be enhanced by a corresponding
increase of the running coupling. We show that this
leads to a faster and improved equilibration. This
improvement is more important for the fermions 
compensating partially for their weaker interactions and 
slower equilibration. \hfill {\scriptsize WU-B 97/30}
\end{abstract}

\section{Introduction}
\label{sec:intro}

High energy nuclear collisions have become a subject of
great interest in the last decade or so, bolstered 
by the experiments at Brookhaven AGS and CERN SPS
and culminating in the constructions of the soon-to-be running
larger higher energy colliders like RHIC and LHC. One
attempts in these experiments to create and study
deconfined matter. An obvious question is how to 
identify matter in the new phase. This is done by the
various proposed particle signatures such as $J/\j$
suppression, strangeness enhancement and excess at low mass 
dileptons. A question of similar importance is the evolution 
of matter in the new state which influences directly
these signatures produced during the deconfined phase
and indirectly in modifying the initial conditions of
the subsequent hydrodynamic expansion. Final hadron
yields are closely related to the generated entropy
which is mainly produced in the initial parton creation
and subsequent parton interactions. This precisely depends
on the time evolution. We study this in order to gain
a better understanding of the evolution and the
equilibration process. 

The main questions of equilibration is whether thermalization
is fast, if parton chemical equilibration can be completed
before the start of the phase transition and the possible
duration of the deconfined phase. Whereas approximate
thermalization should be achieved in the first few fm/c
as shown by models like parton cascade \cite{geig&mull,geig},
completion of chemical equilibration is impossible
for all present existing models 
\cite{geig&kap,biro&etal,wong1,wong2}. As to the duration 
of the parton phase, it depends on various factors. 
Actually the answers to all three questions depend on
the inherent uncertainties associated with these studies.
They are the initial conditions, infrared cutoff parameter
and the value of the coupling used. One can raise the 
question as to whether some of these can be exploited
to determine better or improve the equilibration
process such as parton chemical equilibration. 
To change the initial conditions to get better answers
is too arbitrary and irrelevant since one can vary no 
more than the energy/nucleon and centrality of the collision
in the experiments. Infrared cutoff, which is usual in 
perturbative QCD is actually not needed in a deconfined
medium because of the screening effect of the latter. 
Finally the third possibility is the coupling. One usually
uses a value of $\a_s =0.3$ for an average momentum
transfer of around 2.0 GeV. However, during the time evolution,
the parton energies drop considerably due to the longitudinal
expansion and particle creation. Therefore so will the
average momentum transfer drops in time and the coupling, which 
is treated as a kind of average value here, cannot stay
at a constant value. The exploitation of this effect will
be the main subject of this talk.

\section{Strong Coupling Improvements on Equilibration}
\label{sec:impr}

Our method for this investigation is based on solving
Boltzmann equation with the relaxation time approximation
and explicit construction from perturbative QCD for
the collision terms. All binary collisions and 
2-to-3 gluon multiplication as well as the reverse 
process are included at the 
tree level \cite{wong1,wong2}. In accordance with
our discussion in the introduction, we study
the effect of the coupling on the equilibration. This is
done by comparing results with various fixed values for the
coupling and a time varying coupling obtained by using
the one-loop running $\a_s$ formula and evaluating the 
coupling at the scale, at any time, of the average parton
energy. The reason being that the average momentum
transfer in parton scatterings should be of the order
of the average parton energy. This latter approach allows
the system to determine its interaction strength and
removes an otherwise free parameter, whose 
value was usually chosen with perturbative QCD in mind. 
This will be an advantage as we will see presently
because the results on equilibration do depend on the value
of $\a_s$ used \cite{wong3}. We use the initial
conditions from HIJING for our time evolution 
after allowing for the resulting parton gas to free
stream to an isotropic distribution \cite{biro&etal}.

\bfi
\centerline{
\hbox{
\psfig{file=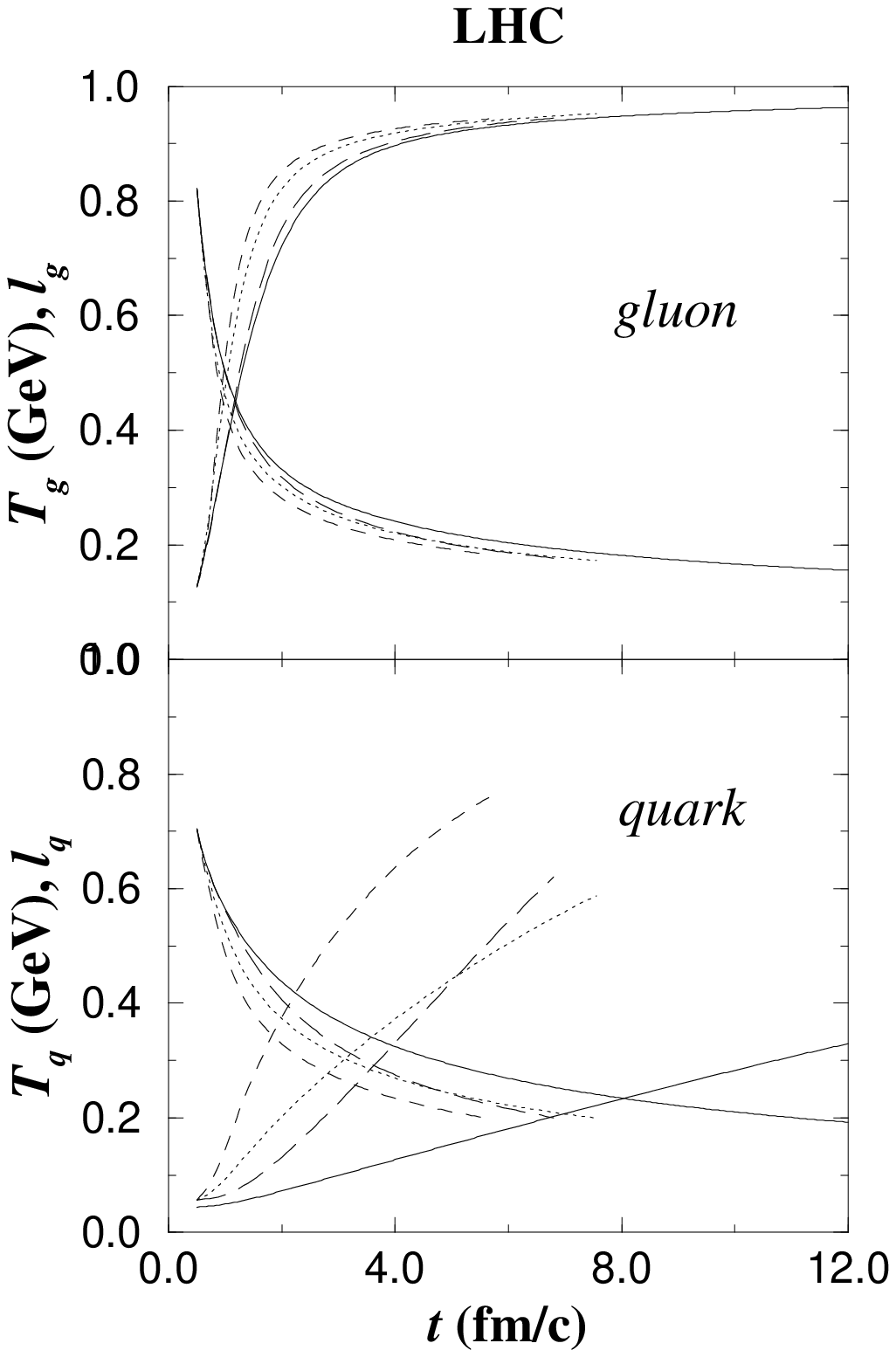,width=1.610in}
\psfig{file=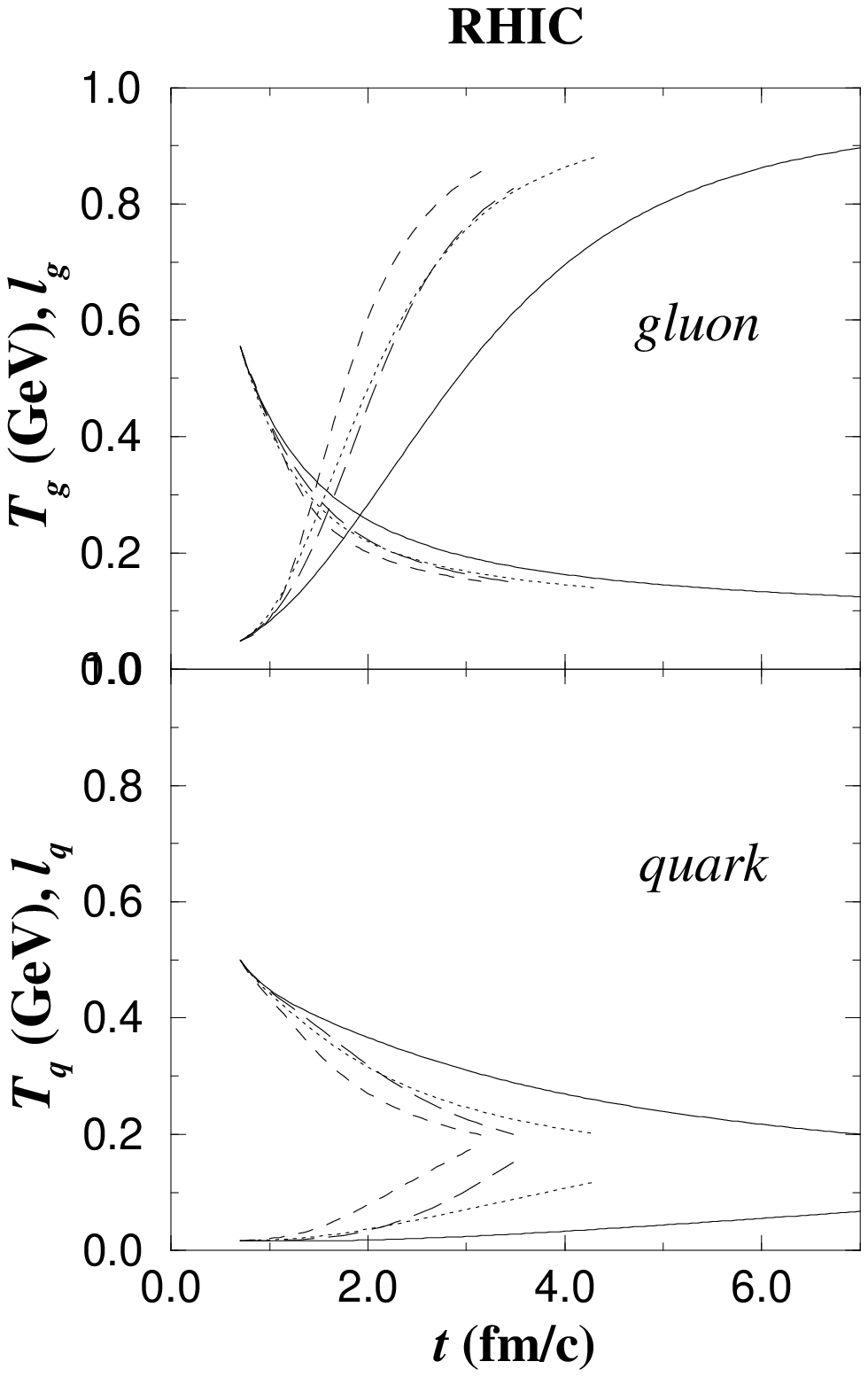,width=1.610in}
}}
\caption{Time evolution of parton fugacities and temperature.
Solid, dotted, dashed and long dashed are for $\a_s=$ 0.3,
0.5, 0.8 and $\a_s^v$, respectively.}
\label{fug&tem}
\efi

To examine the state of the equilibration, we plotted
in \fref{fug&tem}, the quark and gluon fugacities and 
their temperature estimates as a function 
of time and in \fref{pres} the pressure ratios, which 
is a check of kinetic equilibration. 
The various curves are for $\a_s =$ 0.3, 0.5, 0.8
and for the time evolving coupling $\a_s^v$. 
As can be seen in \fref{fug&tem}, larger coupling 
leads to faster chemical equilibration. The fugacities
for both gluons and quarks at $\a_s=$ 0.8 rise faster than
those at $\a_s=$ 0.5, which in turn, are faster than
those at $\a_s=$ 0.3. For gluon, the fugacities at around
4.0 fm/c at LHC and at around 6.0 fm/c at RHIC in all cases
are already reasonably close to 1.0, whereas those
for quarks vary widely with the value of $\a_s$. 
Larger $\a_s$ gives much better results for quarks.
The curves for $\a_s^v$ tend to move in time across those 
with fixed $\a_s$ because of the drop in the average
parton energies due to the expansion and parton creation
which lead to a decrease of the average momentum exchange 
and hence an increase of $\a_s^v$. The resulting 
curves then tend to move from curves of smaller 
to larger $\a_s$. Thus if we use $\a_s^v$, which seems
to be a more consistent choice, parton chemical
equilibration will be faster and improved significantly
in the case of quarks. For gluon, it is hard to get 
any obvious improvement when chemical equilibration
is already very good near the end of the time evolution.

In the case of kinetic equilibration, something similar
happens. In \fref{pres}, the ratio of longitudinal
to transverse pressure as well as one third of the 
energy density to transverse pressure are plotted. 
When the momentum distribution is isotropic as in a 
thermalized system, these ratios are 1.0. This is the
case at the start of the evolution because of our
momentarily thermalized initial conditions. As the 
evolution starts, the expansion pushes the plasma
out of equilibrium and isotropy is lost. The ratios
are seen to deviate from 1.0. At some point in time,
net interactions become fast enough in response to
the expansion and bring the plasma back towards
equilibrium so the curves make a turn and rise
again towards isotropy. 

\bfi
\centerline{
\hbox{
\psfig{file=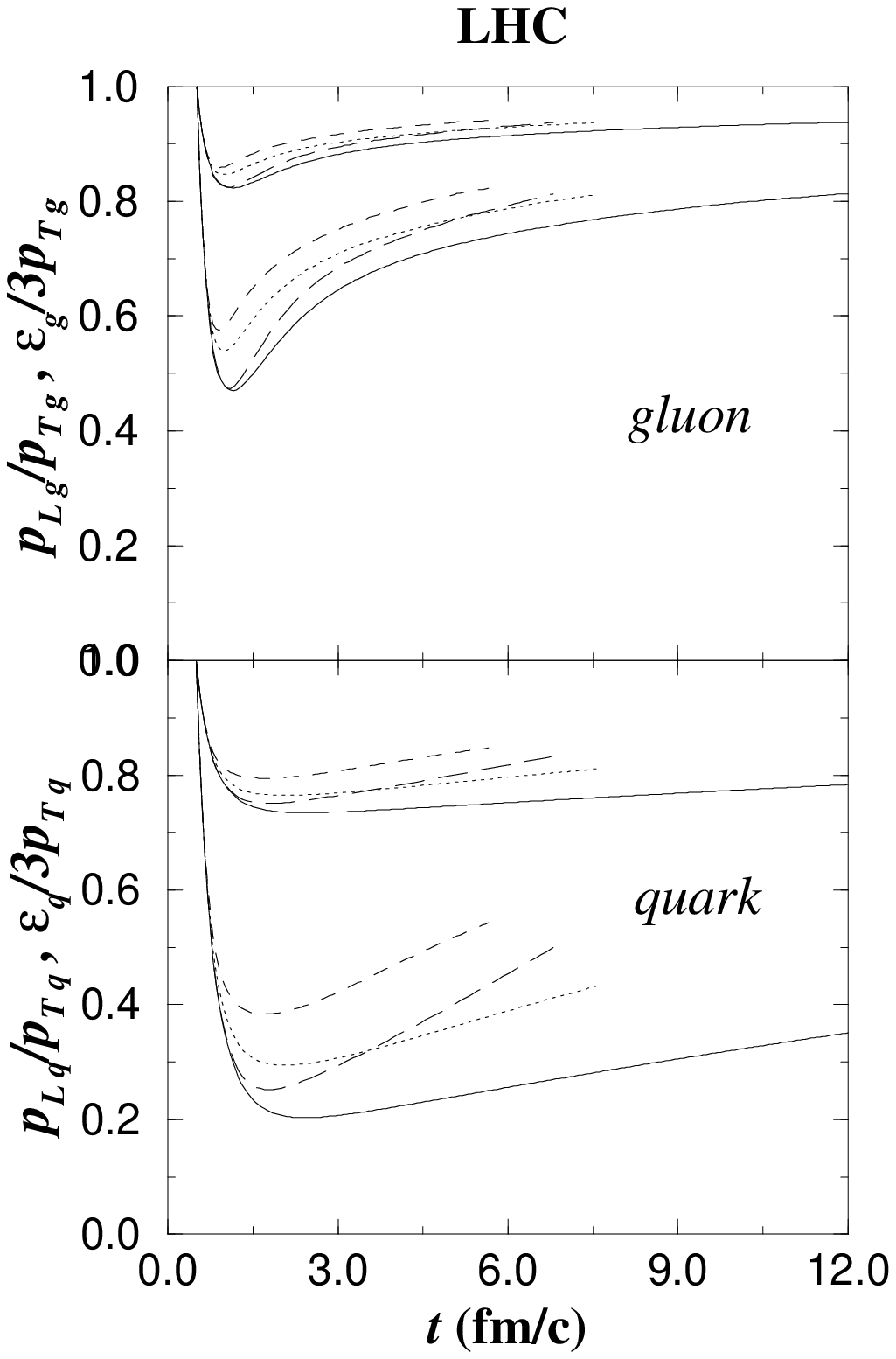,width=1.610in}
\psfig{file=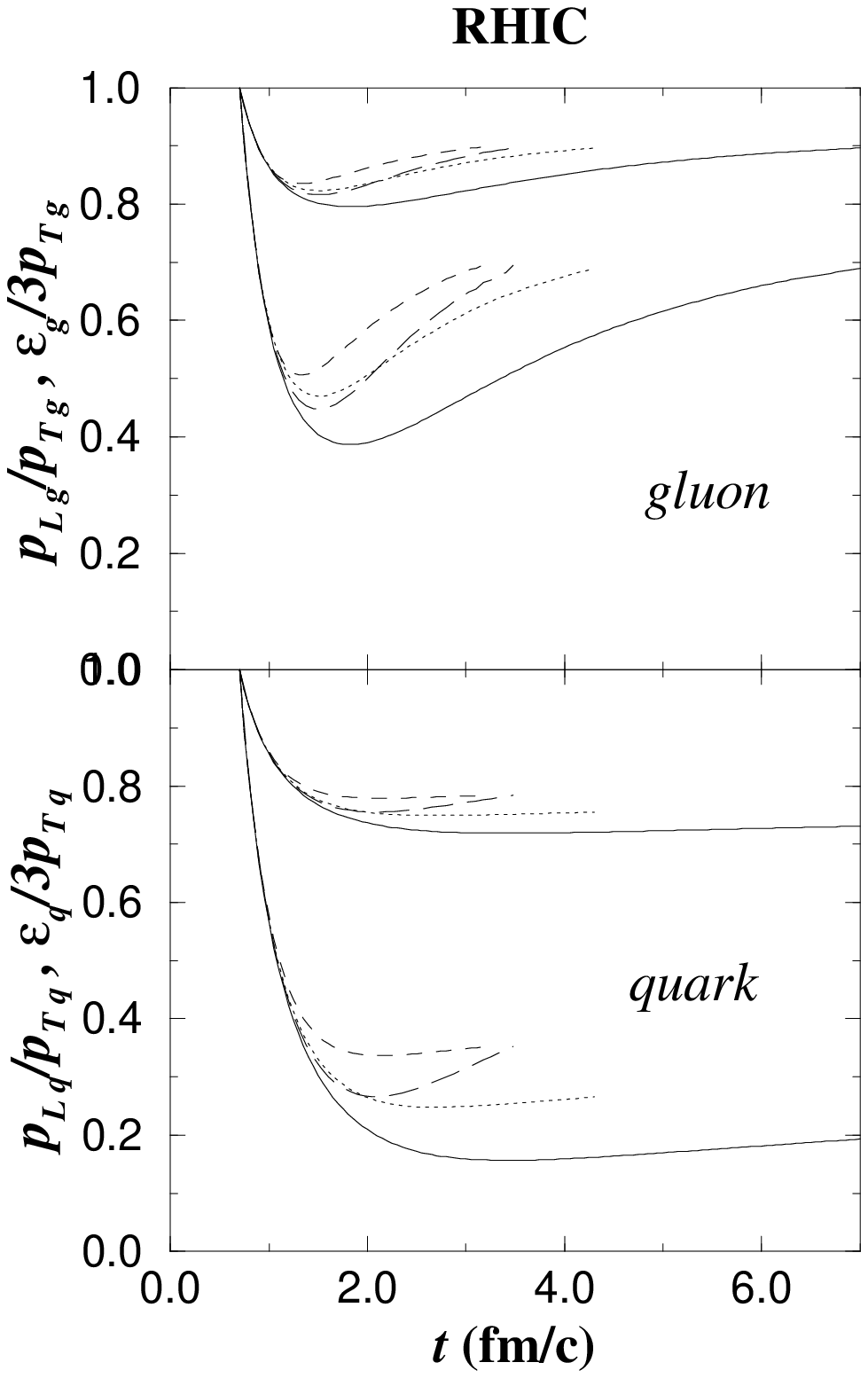,width=1.610in}
}}
\caption{Time evolution of the pressure ratios (bottom
set of 4 curves in each plot)
and energy
to pressure ratios (upper set of 4 curves)
. The assignment of the values of $\a_s$
to the curves are the same as in \fref{fug&tem}.}
\label{pres}
\efi

In the initial stage, there is not much difference
between cases with different values of $\a_s$ because
this is the expansion dominated phase. Larger $\a_s$,
however, causes the interactions to take over sooner
and achieve at the end a better degree of kinetic
equilibration both for quarks and for gluons. Like
chemical equilibration, only quarks but not gluons 
show considerable improvement in the end degree of 
kinetic equilibration. Again, $\a_s^v$ is 
better equilibrated than that of the $\a_s=$ 0.3 
case due to the increasing $\a_s$. 

These improvements are, however, accompanied by a 
reduction of the duration of the parton phase 
resulting from more rapid cooling (\fref{fug&tem}). 
This comes about because of enhanced gluon 
multiplication and conversion into quark-antiquark 
pairs while the quarks have to expand against a 
stronger pressure. So a consistent $\a_s=\a_s^v$ 
speeds up and improves equilibration for the partons
but only at the expense of the duration of the
deconfined phase. 

In summary, because different values of the coupling
have rather large effects on equilibration,
we argued that a time evolving averaged coupling, 
which reflects the decrease of the
average parton energies of the system, should
be used in studying the time evolution in high 
energy nuclear collisions. This more consistent 
approach enhances the equilibration
process both in speed and in bringing the system
closer to full equilibration.

%


\begin{thebibliography}{99}

\bibitem{geig&mull}K. Geiger and B. M\"uller, 
\Journal{\NPB}{369}{600}{1991}.

\bibitem{geig}K. Geiger, \Journal{\PRD}{46}{4965, 4986}{1992}.

\bibitem{geig&kap}K. Geiger and J.I. Kapusta, 
\Journal{\PRD}{47}{4905}{1993}.

\bibitem{biro&etal}T.S. Bir\'o, E. van Doorn, B. M\"uller, 
M.H. Thoma and X.N. Wang, \Journal{\PRC}{48}{1275}{1993}.

\bibitem{wong1}S.M.H. Wong, \Journal{\NPA}{607}{442}{1996}.

\bibitem{wong2}S.M.H. Wong, \Journal{\PRC}{54}{2588}{1996}.

\bibitem{wong3}S.M.H. Wong, \Journal{\PRC}{56}{1075}{1997}.

\end{thebibliography}
\end{document}